# NCore: Architecture and Implementation of a Flexible, Collaborative Digital Library

Dean B. Krafft
Cornell University
301 College Ave.
Ithaca, NY 14850
607-255-9214
dean@cs.cornell.edu

Aaron Birkland
Cornell University
301 College Ave.
Ithaca, NY 14850
607-254-5587
birkland@cs.cornell.edu

Ellen J. Cramer
Cornell University
301 College Ave.
Ithaca, NY 14850
607-254-8952
elly@cs.cornell.edu

## ABSTRACT
NCore is an open source architecture and software platform for creating flexible, collaborative digital libraries. NCore was developed by the National Science Digital Library (NSDL) project, and it serves as the central technical infrastructure for NSDL. NCore consists of a central Fedora-based digital repository, a specific data model, an API, and a set of backend services and frontend tools that create a new model for collaborative, contributory digital libraries. This paper describes NCore, presents and analyzes its architecture, tools and services; and reports on the experience of NSDL in building and operating a major digital library on it over the past year and the experience of the Digital Library for Earth Systems Education in porting their existing digital library and tools to the NCore platform.

## Categories and Subject Descriptors
H.3.7 [**Information Systems**]: Digital Libraries – *Collection, Dissemination, Systems Issues, User Issues.*

## 1. INTRODUCTION
The National Science Digital Library (NSDL) project [33] was created by the National Science Foundation "to provide organized access to high quality resources and tools that support innovations in teaching and learning at all levels of science, technology, engineering, and mathematics education." The NSDL Core Integration team at Cornell University designs and implements the technical infrastructure and tools for the library. Its mission is both to create the best possible library for NSDL and to push the frontiers and capabilities of digital library technology.

As part of that mission, the Cornell team has created a new, open-source, digital library platform called NCore (for NSDL Core). This platform consists of a central repository, based on Fedora[19], a data model and API that define the structure of the library in the repository, and a growing suite of library tools and services that mediate among users, information providers, and the central repository. Since January 2007, NCore has supported the production library activities of NSDL.

While the initial application of NCore is the implementation of NSDL, the platform itself is not specific to NSDL or to STEM education. Instead, it is an architecture and software ecosystem that can support digital library and digital repository needs ranging from cultural heritage materials in the arts and humanities, to scholarly communication and collaboration, to education at every level and in every discipline. Work has already begun on using the open-source release of NCore to catalog and manage the teacher training resources at a major urban public school system and to serve as the central repository and digital library platform for an alliance of eleven major research libraries.

This paper first presents the NCore architecture, focusing on developments since the initial description of the architecture for NSDL 2.0 appeared in [16]. In particular, we provide a detailed explanation and analysis of the use of *aggregations* in NCore, which is one of the keys to the flexibility of the platform. This is followed by a brief description of the back-end services integrated with NCore and a more thorough explanation of the strategy for end-user and community tools and the current production examples. We then present our experience in implementing the Digital Library for Earth Systems Education (DLESE)[1] in NCore, together with a major new NCore platform tool that resulted from this effort. The paper concludes with a brief plan for future work and public release of the open-source code and a summary of what has been accomplished in this effort

## 2. RELATED WORK
This paper builds on extensive work over the past seven years in creating NSDL. Work on the first version of the NSDL architecture, a metadata-based union-catalog paradigm, was described in [15], and a discussion of the design and motivation for the second major version of the NSDL architecture, NSDL 2.0, from which NCore derives, is presented in [16, 18]. Earlier related work on annotation systems, resource linking, and the importance of context for learning is extensively discussed and cited in [16] and will not be repeated here. Earlier work on the role of collections and aggregations in digital libraries is cited extensively in the section below on organizing the repository.

There is a large body of previous work on digital library platforms and the closely related area of institutional repository platforms. Significant open-source digital library platforms in wide production use include Fedora[19], Greenstone[31], DSpace[30], and EPrints[23]. Compared to the latter three, by building on top of Fedora, NCore inherits many of Fedora's key advantages: an open architecture and data model; a highly flexible architecture of relationships among digital objects in the model; and the easy ability to extend the repository, metadata, relationships, and content types. Compared to the base Fedora system, a middleware package that requires extensive development to create an end-user accessible tool, NCore provides a specific data model, organizing relationships, and a wide suite of extensible tools and services. Like Fez [13], NCore is built on Fedora, but it is much more of an extensible and integrated platform of digital library tools than Fez, which is designed as a digital repository management and workflow system.

---

[1] http://www.dlese.org

The NCore end-user tools exist at the intersection of education, digital libraries and Web 2.0. Frumkin [7] discusses using wikis in digital libraries, and speculates on applying a wiki as a digital library annotation tool. Milson and Krowne [24] describe re-conceptualizing the digital library as a dynamic, commons-based peer production knowledge environment, and they examine this concept using PlanetMath (http://planetmath.org) and Noösphere [14], a wiki-like collaborative environment, in an instructional setting. Downes [6] describes a vision of *e-learning* where "online learning software ceases to be a type of content-consumption tool, where learning is 'delivered', and becomes more like a content-authoring tool, where learning is created", a vision in close harmony with that of NCore.

## 3. NCORE: THE CENTRAL CORE

At the heart of the NCore platform lies the Fedora-based repository, the data model and digital objects that define the content of the library, and the relationships that organize the materials and provide both structure and context. Real life, real resources and real information are never neat and hierarchical. Instead they form a complex web of relationships and bits of information with varying degrees of certainty. NCore is designed both to capture and represent this chaotic reality, and to make it accessible to users and other services in ways that enable discovery, usability, and understanding.

### 3.1 The Repository and Data Model

A full description of the initial repository architecture of NCore can be found in [16, 17], but we will briefly review the key concepts here. The rest of section 3 will discuss changes to the architecture and implementation as a result of two years of development and production experience since the initial report.

The library is a set of digital objects and relationships, stored in the Fedora-based NSDL Data Repository (NDR) and accessed through a REST-based web services API (see section 4.1). The primary digital objects are: a *resource object* that either contains or specifies content (e.g. a web page, a journal article, a dataset); a *metadata object* that contains structured statements about a resource (e.g. author, title, audience); an *aggregation object* that collects resources and other aggregations together in a set; a *metadata provider object*, a special type of aggregation object that aggregates and provides provenance information for metadata objects, and an *agent object* that specifies the source for metadata statements and the selector for aggregations. In addition to the structural relationships among these objects implied by the descriptions above, NCore also supports other arbitrary typed relationships among the objects.

Building on the underlying Fedora relationship architecture[19], NCore also serves as a *semantic digital library*, with the capability to support sophisticated, arbitrary queries over the relationship structure of the repository. At a technical level, these relationships are expressed in the repository as Semantic Web-style RDF triples[12]. In the current NSDL implementation, they are held in an MPTStore-based triplestore (http://mptstore.sourceforge.net), supporting simple queries with complex query support (a subset of SPARQL) in development. These relationships can also be held in a compatible Mulgara (http://mulgara.org) triplestore, which allows complex iTQL queries over the triples[32].

### 3.2 Organizing the Library with Aggregations

Organization and characterization of content have always been essential functions of traditional and digital libraries alike. Digital libraries in particular afford inexpensive means of grouping or otherwise co-locating resources into discrete "collections," which may be browsed, searched or otherwise manipulated as a whole. In many digital libraries, including union-catalogs such as NSDL 1.0, these aggregations are quasi-static: growing in number and size over time according to a given curation policy, yet any given resource, once in a collection, is unlikely to find itself re-purposed into another.

As digital libraries have evolved, many have identified user-contributed content, personalization, and re-purposing of content as essential value-add features of "Next Generation" digital libraries. In response, there has been much research exploring the principles and technologies behind this functionality. For example, Smeaton and Callan [29] describe the characteristics of personalization, recommendation, and social aspects in next generation digital libraries, while [1, 26] describe an implementation of personalized recommender services in the CYCLADES digital library environment. Crane et al. [5] emphasize the need for decentralized community contributions to library content such as annotations in Wiki pages. Contributions by faculty and researchers maintaining their own collections and information spaces in a digital library are explored in [3] as part of the ADEPT project.

In general, the research has emphasized the need for next-generation digital libraries to contextualize their resources with contributions from disparate sources. In such an environment, data surrounding a resource, such as subject metadata or membership in an aggregation, does not purely originate from a single cohesive and consistent curation policy, but from a variety of independent agents with their own motivations. NSDL, in its conception of NCore, has realized that from this perspective, there is no single authoritative metadata record for any given item in the library. Instead, an item is described by multiple records, each describing the resource from the perspective of a particular information provider who is contributing to the library. Likewise, a single resource may be a member of arbitrarily many aggregations or collections, being organized by the same set of independent agents with differing motives.

Faced with the prospect of managing this multi-sourced and potentially user-contributed context, the topics of access and control become particularly relevant. How can a library curator retain editorial control over which user-contributed content is considered to be "in" the library's public face? How can this content be incorporated into library services in a way that provides additional value rather than additional noise? In fact, many challenges of next generation digital libraries can be framed in terms of management and interpretation of aggregations. Thus, there is a strong case for designing digital library infrastructures with aggregations as first-class objects. The NSDL has adopted this approach in its design of NCore, where aggregations occupy a central role in representing and mediating context within the repository.

## 3.3 Defining and Characterizing Aggregations

The word "collection" as it applies to digital libraries can seem familiar, ambiguous, and loaded at the same time. There is much literature in which the term's meaning is assumed to be understood, yet in those instances where a "collection" is defined, it is not always defined consistently, nor do these definitions always share the same characteristics [10, 20, 25].

In the most basic sense, a collection is just an arbitrary bag of digital objects, and it is up to some sort of explicit or implicit collection metadata to provide specific semantics for the collection [10]. Traditionally, however, the term "collection" has been understood to imply a certain degree of ownership or control [20]. Indeed, it seems that many view traditional digital library collections as a partitioning of the digital library resources. Geisler et al. [8] signify a break from that view in distinguishing "virtual collections" as being created from existing library resources, yet existing alongside the "real" collections that form a partition. Meghini and Spyratos[22] concur, labeling the same dichotomy as physical vs. virtual, and further classify virtual collections as static or dynamic. Static virtual collections are taken to imply a long-lasting assembly of resources for a particular purpose oriented towards some community, whereas dynamic are taken to be user-created aggregations that support a particular task or reflect an individual's view of current library contents for some duration of time.

At this point, it makes sense to consider the distinction between an aggregation and a collection. As previously noted, the term "collection" in a digital library sense implies a certain degree of semantic meaning or intent. "Aggregation", on the other hand, tends to imply merely an assembly of items and nothing more. For the purposes of this paper, an aggregation is defined as a named set of digital library objects, where digital library objects may be primary digital content (resources), metadata, aggregations, or agents. In this light, a collection is an instance of an aggregation that carries with it some specific semantics.

Through the experience of developing the NCore architecture, we have come to appreciate aggregations as one of the fundamental building blocks for various structures found in the library, collections being only one example. As such, we have identified some relevant characteristics to successfully engineering working structures out of aggregations:

- Aggregations may be described by metadata;
- Aggregations have heterogeneous membership unless otherwise enforced;
- Membership in an aggregation is controlled by some entity (agent, in NCore terms);
- Any object may be a member of arbitrarily many aggregations.

The following sections will demonstrate the utility of each of these characteristics and present these characteristics in light of the experience of NSDL and other efforts.

## 3.4 Structure of Aggregations

Aggregations are the building blocks for many structures that occur in digital libraries. Here, we explore some of these structures and their characteristics as applicable to a collaborative digital library such as the NSDL.

### 3.4.1 Nested Aggregations

Nested aggregations, as the name implies, are structures formed by containing aggregations within aggregations. The ability for aggregations to nest is a prerequisite for implementing sub-collections or achieving any sort of hierarchical organization of library resources. The original incarnation of NSDL, perhaps typical of union-catalogs of its era, did not have general-purpose aggregations. Its collections were single-level, containing item-level resources only. In that era, all data in the library came through OAI harvests, and each OAI data source was mapped to an NSDL collection. In order to support the goals of NCore and to permit applications to contribute, organize, and repurpose resources in the library, it became clear that a single-purpose, one-level hierarchy over NSDL resources was insufficient.

### 3.4.2 Multiple categorization

Although nested aggregations may be used to create hierarchical structure, nested aggregations do not imply a hierarchical structure. Indeed, in an environment such as NSDL, where many independent agents have the ability to create new aggregations, the resulting structure is far from hierarchical. A hierarchy implies that each member has exactly one parent. In order to support multiple agents creating their own orthogonal organizational structures across a shared set of resources, some resources and aggregations must be members of more than one aggregation.

There is also strong case that allowing objects to be a member of multiple aggregations is a powerful tool to hand to users. Karger and Quan [11] argue that multiple-categorization is more valuable to users organizing data than are hierarchies, and find that users are generally "less inhibited" in doing so. Indeed, with multiple-categorization, assigning a resource to a particular aggregation does not come at the cost of removing it from another.

### 3.4.3 Complex objects

Complex objects are single entities that are composed of multiple parts, each of which is an entity in and of itself. In order to support complex objects in a digital library, it is necessary to demarcate the "boundary" around a set of resources and manipulate that composite as a first-class object. Buchanan et al. [4] describe these as composite aggregations, and note that they represent a particularly difficult class of aggregation that is problematic in the few digital library systems that support them.

The importance of aggregations in defining complex objects is recognized not only in the digital library context as in [4], but also plays an important role in efforts such as OAI-ORE (http://openarchives.org/ore/) that focus on exchange and interoperability. With that in mind, complex objects may currently be represented in the NCore model as an aggregation containing the constituent members on an ad-hoc basis. At present, the NSDL is awaiting the formal OAI-ORE specification and related discussion to inform further development of complex object support. While it is certain that complex objects will be based on aggregations, to truly support them in an interoperable fashion is likely to require representing additional semantics on top of the base NCore model, perhaps in the form of specific object properties, relationships, or metadata.

## 3.5 Semantics of Aggregations

There is overwhelming consensus on the importance of metadata to describe the semantics of collections [2, 8, 10]. Since aggregations themselves are devoid in semantics (but rich in context), it is apparent that the ability to describe aggregations with metadata is necessary. Meghini and Spyratos[4] characterize aggregations in terms of extension (the set of objects within it) and intension (the meaning of the aggregation, as differentiates it from others and specifies a homogeneity criterion for the resources within it). In that sense, in the NCore model, aggregations themselves exclusively represent extension, while aggregation (collection) metadata statements exclusively represent intension.

While it is important to have the ability to describe the intension of an aggregation, we have found that it is equally important not to require it, nor to require a particular standard of quality or completeness. In a sense, this echoes the sentiment of [8], in that for certain tasks, such as organization of resources as encountered in personal information management, ease of use is the dominant requirement. Indeed, any description of an aggregation a user provides is likely to be very different from metadata describing a curated collection. Folksonomic tagging[9] is perhaps an appropriate example of a form of lightweight metadata that describes an aggregation in a meaningful way to a user.

### 3.5.1 Property/membership duality

There is more than one way to classify a resource. There exists an uncomfortable duality between aggregations and metadata properties when either membership in an aggregation or a metadata property are able to achieve the same goal of classification[8, 25]. For example, is it better create an aggregation of resources that conform to a particular educational standard, or is it better to create metadata for each resource saying so directly?

In answering the question of duality, it is worthwhile to consider some of the implications of first-class aggregations. With first-class aggregations, membership data is not expressed as a field or property in descriptive metadata. As Buchanan et al. [4] points out, simply encoding membership as a metadata field is ineffective: it is prone to errors in consistency, and it does not scale. With first-class aggregations in NCore, aggregation membership has the advantage of being subject to consistency constraints (e.g. it is impossible to claim membership in an aggregation that does not exist), security constraints (e.g. nobody but an owner of an aggregation and any specifically authorized parties may change its membership), and being handled in a consistent and predictable manner by library user interfaces.

### 3.5.2 Inheritance

Given nested aggregations, there is the question of whether properties of the parent also apply to the child. As far as descriptive metadata, there doesn't seem to be a universal consensus[8]. For aggregation membership, however, there is no ambiguity. Children of nested aggregations are defined to be related to their ancestors by transitive membership. NCore services such as search make use of this definition, and allow selection of all resources that are "under" (i.e. related via direct or transitive membership) a given aggregation. While all the implications of this are out of the scope of this paper, the concept of membership inheritance is important for using aggregations to demarcate "areas" in the repository in a scalable fashion by building them from nested aggregations rather than individually.

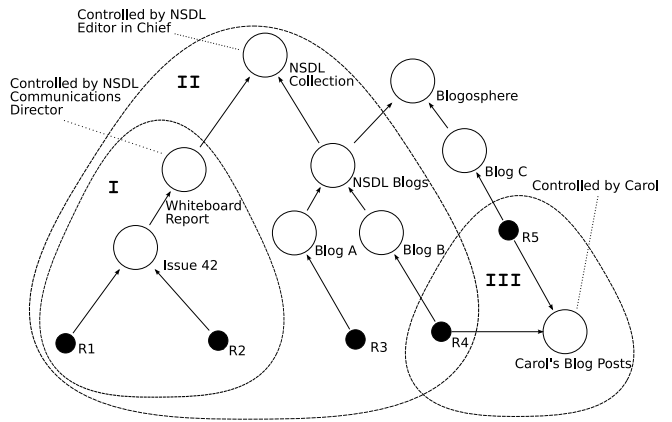

**Figure 1 – Diagram of NCore repository with resources (filled circles), aggregations (empty circles), and directional membership arrows.**

Figure 1 illustrates a forest of nested aggregations in an NCore repository. For example, Region I represents part of the content and structure of the NSDL "Whiteboard Report" publication. Individual articles R1 and R2 are aggregated into Issue 42, which in turn is a member of the overall "Whiteboard Report" aggregation. Considering membership as a transitive relation, each of R1, R2, and Issue 42 are members of the "Whiteboard Report" aggregation, and also members of the top-level "NSDL Collection".

## 3.6 The Library Perspective on Aggregations

Given first-class aggregations as a primary means for organizing data, there are several ways that digital libraries may use these aggregations to provide or enhance digital library services.

Browsing resources and collections is an important service for a digital library portal to provide. Browsing complements search as an effective tool for resource discovery, particularly for resources that are non-textual. An important aspect to browsing is co-location of resources. Traditional library schemes support co-location of materials by design of classification systems, but digital libraries can be more effective in this regard since they can support co-locating resources by many different criteria [8]—and it is cheap to do so. Aggregations are a means by which resources may be co-located for the purpose of browsing. By incorporating aggregations contributed by library users, a library increases the degree of co-location and enhances serendipity by browsing as a plausible means for discovering resources.

A number of research initiatives have demonstrated that user-contributed aggregations may be used for providing personalization or recommendation services in digital libraries [8, 22]. In general, such services rely on developing similarity metrics between aggregations or between items and aggregations. Renda et al.[26], for example, provide an algorithm for calculating the "centroid" of the terms found in the documents within an aggregation, and are able to compare this with the terms found in any given document. The degree of match is used to determine if a particular resource is similar to the resources within the aggregation, and thus a candidate for recommendation.

NSDL has not yet implemented any such recommender services, but has identified this as an area for future research and potential implementation. In encouraging the creation and use of aggregations in NCore and its related tools, and by soliciting user-provided content, NSDL has ensured that the platform fully supports these potential extensions.

## 3.7 Motivating Users to Create Aggregations

As mentioned in the previous section, user-created aggregations can add significant value to the library, leveraging the collective intelligence of the users to enhance services for browsing and recommendation, among others. But how do these user-contributed aggregations make it into the repository? Why would a user want to organize library resources into aggregations in the first place? What's in it for the user?

Perhaps the most interesting mechanism for adding user-provided content and organization to the library is as a by-product of other user interactions with library tools. As will be described in section 5, NCore includes familiar content-creation tools, such as WordPress, for blogging, and Mediawiki, for wiki creation, that integrate their content and relationships into the NCore repository. Thus, the action of writing/editing a wiki page or creating a blog post results in content and referenced resources being incorporated into the NCore repository. These tools aggregate user contributions by source, so, for example all a user's blog posts may fall into an aggregation, or the resources mentioned in a blog post may be aggregated together, as well as by the structure imposed by the particular tool, so that all posts to a specific blog or category may form an aggregation.

Personal information management is another means by which user-contributed data may find its way into the library. Borgman et al.[3] found that personal digital libraries were not only useful for geography faculty to collect and organize resources for their teaching or research, but also in providing resources and context to the library.

NSDL is currently investigating how best to incorporate personal bookmarking/tagging systems, such as Connotea, del.icio.us, and Technorati, into NCore. In such a system it would be easy to create an aggregation composed of all the resources bookmarked by a single user, or all those tagged with a particular folksonomic tag.

Application developers and contributors to the library can also benefit from creating aggregations in the library. Doing so can expose the aggregation in search and browse interfaces. Aggregations can also be used to "brand" resources as part of a particular collection. Several NCore tools (see section 5) can be used to create and manage such collections in the repository.

Aggregations in the repository can also be exposed to outside services as an RSS or ATOM feed. By managing the resources in an aggregation, NCore affords a simple mechanism for controlling mashups with applications that consume these feeds.

## 3.8 Managing Aggregations

So far, we have shown that many digital library structures such as collections, complex objects, and personal information spaces may be defined in terms of aggregations of resources with the characteristics detailed in section 3.3. We have also shown that these aggregations are useful to users and library-oriented applications for contributing and re-purposing library content. The content and organization contributed by these users and applications via aggregations may be incorporated by the library at will to support or enhance library services such as multi-faceted browsing, presenting the context around a resource, or the creation of personalization or recommendation services. As a result, there is a strong case for digital libraries to expose aggregations as first-class read/write objects to users and curators alike.

As first-class objects, membership in an aggregation is separate from the metadata properties that may describe a resource in the library. Access to an aggregation's members or parents can be achieved in a uniform fashion, and may be subject to universal rules regarding consistency and permissions. The NCore model and API implements all these characteristics in the context of a Fedora repository. It provides a read/write API to the underlying objects, specifically treats aggregations as first-class objects with requisite functions to manipulate them, and provides a security and referential integrity model for aggregation membership.

In conjunction with a consistency and permissions model, such as that provided by NCore, aggregations may be used to mediate the contributions of individual agents in a repository and enable building a cohesive library from these disparate pieces. As mentioned earlier in section 3.5.2, aggregations may be used to define the boundaries around "areas" in a repository. For this purpose, recall that aggregation membership is considered to be a transitive property. Aggregations, then, may be used to define the boundaries of a digital library itself within a repository.

For example, one may consider a library to be defined as composed of the objects specifically *in* the library and those specifically considered *not in* the library, where membership in both sets implies *not in*. Two aggregations, combined with transitive membership, can realistically be expected to completely represent the boundaries of a digital library in terms of the resources within it. In a more general sense, aggregations may be used for defining arbitrary "views" of content within a repository.

NSDL, for example, may be defined as an aggregation representing the extent of the library. Within this aggregation are the aggregations of all the collections that are considered to be part of NSDL. Implicitly, these collection's members are also considered to be part of NSDL by transitive membership. This implicit membership is important, since it eliminates the need for every item to be explicitly added to the NSDL library aggregation. Without it, such definition would not be scalable or maintainable.

Referring again to Figure 1, the entirety of the NSDL library is represented as the area underneath the "NSDL Collection" aggregation, denoted as region II. As is evident, there are only two direct members of the NSDL aggregation—all items underneath these two are members of the "NSDL Collection" aggregation due to the transitive nature of membership.

Continuing with the example of NSDL, transitive aggregation membership has interesting implications for control over the exact shape of the library boundaries. Since access to an aggregation is controlled by policy, the decision as to whether to add an aggregation to NSDL can be controlled by the curator of NSDL as a whole, and no malicious agent may change that selection. Nevertheless, a collection contained in NSDL may not be under the control of NSDL. Thus, while NSDL selects a collection as being *in* the library, it does not control the membership of that collection itself. This is a form of delegated authority that arises when one mixes aggregations of different ownerships, and is a motivation for creating an explicit "*not in* NSDL" aggregation where the curation policy for NSDL may not match the curation policies of those collections operating under delegated authority.

As illustrated in Figure 1, the two aggregations that are direct members of the NSDL collection aggregation are controlled by the editor in chief. One of these members is the Whiteboard Report. The Whiteboard Report itself is controlled by a different entity entirely: the NSDL communication director. Likewise, consider region III of the graph. These are blog postings that were created by Carol and are in her aggregation. While she is only interested in demarcating these posts as having originated from her, she does not control the fact that one of them (R4), is a transitive member of the NSDL Collection.

### 3.9 Aggregations of Metadata

Up to this point, we have discussed aggregations containing resources or other aggregations. Since the members of an aggregation are not inherently restricted by type, it is worth mentioning the consequences of aggregating metadata. The NCore model provides a particular kind of aggregation, called a *metadata provider*, that is used for determining the provenance of every metadata statement in the repository. Every metadata object in the repository *must* be a member of exactly one *metadata provider*. Aside from this special but important, case, metadata may be additionally aggregated just like any other object in standard aggregations. To do so plays an important role in defining libraries, as mentioned in section 3.8. Since NCore separates resources from metadata, and any given resource may be described by multiple metadata statements from multiple sources, there must be a mechanism to decide which metadata statements are *in* the library, and which are not.

While NCore allows such aggregations of metadata, fully supporting these to create independent views of the library is dependent on indexing services (see section 4.3). We are currently investigating appropriate index strategies that would fully support filtering search queries by both resource *and/or* metadata aggregation at the same time.

## 4. NCORE: BACK-END SERVICES

A major challenge for NCore was the need to support a highly robust and scalable digital library platform. To support the needs of NSDL, NCore must provide 7x24 operation; high availability and quick recovery; security, authentication and authorization; support for one of the largest Fedora repositories currently in production; and an automated workflow capable of handling over 150K resource updates per month with minimal staff intervention.

### 4.1 The Production NSDL Data Repository

NSDL on the NCore platform is currently in production and accessible to the end user through http://nsdl.org. As of January 21, 2008, the library contained 3.02 million resource objects, 2.3 million metadata objects, 990 aggregation objects, and 816 agents. To support the high availability requirements of NSDL, the production system makes use of a Fedora-level transaction journaling system developed by the Cornell NSDL team. Transactions on the repository are replicated in real time to two "follower" systems, ensuring minimal downtime for all updates and failures.

Web services access to the library is available through the NDR API[2], which uses public/private key authenticated access to allow registered agents to add and update metadata and aggregations. The NDR API is used for all update access to the repository by both the back-end services and the front-end tools. It also supports public read access to the open materials in the repository.

### 4.2 OAI-PMH: Harvest, Ingest, and Serving

A major source of resources and metadata for NSDL is the aggregation of Open Archives Initiative–Protocol for Metadata Harvesting (OAI-PMH) feeds. The process and tools currently used for OAI-PMH harvesting and ingest are described in [15]. However, as part of the development of NCore, a new harvest management system is being designed and implemented. This will integrate with the new NSDL Collection System (described in section 6.2 below) to support automated scheduled harvests from OAI-PMH providers.

The metadata harvesting and ingest process creates an aggregation of all the resources associated with a particular metadata provider, and a separate aggregation of all the metadata objects. These aggregations can overlap with other existing library aggregations, for example when two metadata providers both describe the same web resource. Since an OAI-PMH metadata provider is defined by the organization, the OAI server, and the particular set, arbitrarily granular collections can be created for a single organization or OAI server.

To serve out batch metadata using OAI-PMH, NCore uses the Fedora OAI Provider Service[3]. This supports exposing any metadata format available in the NCore repository in a highly flexible manner, where OAI-PMH sets can be based on NCore collections, aggregations, and relationships. Because it supports batch incremental updates on repository resources and metadata, the OAI Provider Service is used by several internal and external services, including archiving and search.

### 4.3 Search

The NCore platform includes a REST-based search service built on Lucene and Nutch. The service provides fielded search over NSDL's qualified Dublin Core metadata for the resources, as well as the full text of the resource as crawled by Nutch. This fielded search supports the selective propagation of aggregation-level metadata fields down to resource items within an aggregation. The search web service exposes the full power of the Lucene query language to discover resources in the library.

---

[2] http://wiki.nsdl.org/index.php/Community:NCore

[3] http://fedora.info/download/2.2/services/oaiprovider/doc/

The search service can filter resource search results based on aggregation membership, allowing a single search service to support multiple "views" of the library at the resource level. It is also possible to use the search service to obtain metadata-level "views" of the library by including or excluding specific metadata providers and their associated aggregations of metadata (see section 3.9). However, each such view currently requires building a separate search index.

The search index is currently updated nightly using incremental harvest from the repository's OAI provider feed. While satisfactory for OAI harvested collections, the delay is undesirable for resources and relationships created by the new NCore interactive front-end tools. Work is underway to support very fast incremental updates to the search index.

## 5. NCORE: FRONT-END TOOLS

The quality and flexibility of user-facing tools is critical to achieving the goal of creating a *collaborative* digital library. Fortunately, the Web 2.0 phenomenon has unleashed a flood of open-source tools that specifically support user contribution and collaboration, with the goal of building value by harnessing the collective intelligence of the users of the Web.

The NCore development team has sought to leverage existing general open-source Web 2.0 tools (e.g. blogs, wikis) as well as specialized tools (e.g. course management systems, learning module creation tools) by writing simple plug-in extensions that integrate these tools into the NCore platform. By minimizing the development effort required to integrate a tool into NCore, the team has maximized the quality, range and impact of the tools that are being made available.

To support user authentication for the front-end tools, NCore makes use of a highly scalable sign-on system using the Internet2 Shibboleth technology (http://shibboleth.internet2.edu/). In its implementation for NSDL, the primary identity provider for community sign-on is operated by Columbia University as part of NSDL Core Integration. However, the tools and authentication will operate with any appropriate Shibboleth identity provider.

### 5.1 The NSDL.org Web

The primary public channel for access to NSDL and the contents of the NSDL repository is through the web portal at nsdl.org. The site supports several different access mechanisms to NSDL resources and metadata. The search and search results interface provides a number of specialized audience views of all the materials in the repository that have been chosen to be "in" the library. "More info" and "resource page" views of resources provide a complete picture of all the information that is known about a resource: collection membership, metadata statements and relationships to other resources. The "resource page" views are also indexed by Google and other search services.

Other user interface views of the library include browsing by subject, collection, and Science Literacy Maps[4], which allow teachers and students to graphically explore the space of interrelated STEM concepts, associated educational standards and benchmarks, and the library resources related to those concepts and standards.

### 5.2 Expert Voices: Blogging in NSDL

Expert Voices was developed as a collaborative tool to increase community contributions to the library, relate library resources to real-world science events, and provide context for science resources in the library. Expert Voices provides the infrastructure for engaging teachers, scientists, librarians, and students in conversations about STEM topics. As an integrated component of NCore, Expert Voices makes it easy for users to find content from the library, and it allows them to exchange ideas and point each other to useful online materials.

There are a number of models for making use of Expert Voices blogs within NSDL. These include the discovery team model, in which teams of teachers, scientists, and media specialists blog about science discoveries and real-world science applications; the classroom model, where teachers use blogs to create lesson plans for their students, and students then use them for writing and collaboration [27]; the community model, where members of a particular science and education community present news, discuss topics of interest, and promote educational outreach; and the research dissemination model, where a particular research team uses the blog to present ongoing research activities, research results, and links to publications and related work.

Blogging provides a low barrier opportunity for time-constrained teachers to connect to busy scientists. Scientists, in turn, can share their knowledge and zeal through a blog, using it to debate the results of studies or events in real time, organize information, and relate their work to background materials, relevant areas of science, and the real world[28].

Expert Voices has many individual blogs on a variety of topics, designed for various audiences. To help visitors find posts of interest, the home page of the Blogosphere has a section displaying blog titles by audience, another for posts by topic or category, and a section displaying the more recent posts in Expert Voices. Because the system is built on popular blogging software, the basic functionality is familiar to the average blog user. Experienced visitors use their favorite news reader to point to specific blog RSS newsfeeds. There is also a plug-in for email subscription for those not as comfortable with RSS newsfeeds.

Expert Voices is built using a standard, open-source blogging system (WordPress MultiUser[5]) and supports blogging standards, themes, templates, and plug-in functionality. In addition to being able to add and edit blog content, authorized contributors can also add new resources to NSDL, embed links in their blog entries to new and existing NSDL resources, and add metadata to resources, all via custom WordPress plug-ins. These plug-ins utilize publicly available NSDL REST-based web services: the NSDL search service and the NDR API

Expert Voices forms a collection or aggregation, and each blog is an aggregation whose members are individual blog entries. When a blog post is published to the NDR, the blog author can either reference existing NSDL resources within the post, optionally adding new metadata, or they can create new resource entries in the library by adding a reference to the resource together with basic resource metadata (see figure 2). Within the NDR, the blog entry serves as an annotation of the resources it references. It also imposes a human-created inferred relationship among all the resources that it annotates. This information can, in turn, enhance other users' access to library resources. If a user navigates to one resource annotated by a blog entry, other resources that are also annotated by that entry may well be relevant.

---

[4] http://strandmaps.nsdl.org/

[5] http://mu.wordpress.org/

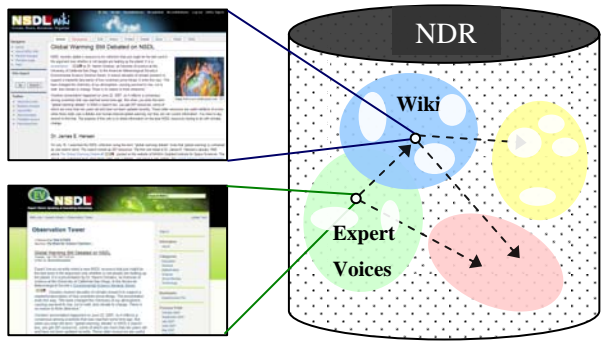

**Figure 2 – Blog posts and wiki pages may be resources in the NDR and have relationships with referenced resources in other aggregations.**

The Expert Voices custom plug-ins for interacting with the library are designed to be as simple as possible to encourage the discovery and contribution of resources. The search popup plug-in allows users to find resources, and it formats the links for insertion into the blog content. The contribution popup plug-in provides online forms for simple metadata submission, and it also creates all the metadata, resource and aggregation relationships in the library. Not all blogs are added to the library, and only trusted users have the authority to publish resources and posts in the library. Feedback from the NSDL partners indicates that the ability to easily add resources using these collaborative tools is a high priority for their users.

## 5.3 The NSDL Wiki

The NSDL Wiki is the second major collaborative tool to be integrated into NSDL. The core MediaWiki software (http://mediawiki.org) is used by millions of Wikipedia users and contributors every day. It provides a familiar functionality of collaborative authoring using a simplified markup language, hyperlinks, and user categories to create and modify wiki pages. In addition to the default wiki functionality, the NSDL Wiki provides the ability to add newly created wiki pages to the NSDL Data Repository as resources with simple structured metadata (see figure 3).

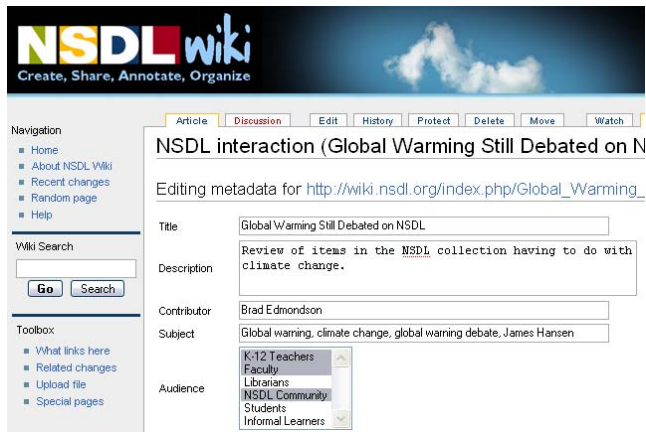

**Figure 3 - Adding a wiki page with its metadata to the NDR using the MediaWiki extension.**

Users or groups can also use the wiki pages to collect and organize NSDL resources for information dissemination or for teaching. A wiki editor can directly reference NSDL resources as well as pages from other wikis or the web. These organizational pages can, in turn, be added back to the library as new aggregations of the resources they reference. The aggregations are then available as part of the library, accessible through nsdl.org, the search service and NDR API, for other users to discover and repurpose.

## 6. IMPLEMENTING DLESE IN NCORE

The Digital Library for Earth Systems Education (DLESE) is a long-standing and successful effort to create a community digital library of geoscience materials [21]. Over the past eight years, in addition to the resources and metadata in the library itself, the project has created a significant and valuable infrastructure of tools, processes, and standards for metadata and collections to support the library.

In 2007, DLESE was challenged to come up with a sustainability model that would free the project from needing to run on dedicated hardware and software systems. To achieve this, the DLESE project and its partners at Digital Learning Sciences decided to implement DLESE on the NCore platform, and to potentially migrate the entire existing library, its processes, services, resources, and metadata, into the NSDL Data Repository. This would allow DLESE to implement their community library model through a standard hosted web site linked to the data, services and tools hosted on the NCore platform by NSDL Core Integration, dispensing with DLESE's dedicated hardware, software, and associated system administration and support.

The effort was successful, demonstrating the flexibility and extensibility of the NCore platform and adapting previously DLESE-specific tools to integrate with NCore.

## 6.1 Modeling DLESE Data in NCore

The DLESE data model consisted of a combination of collection records and associated item records, where each item record was a member of a single collection. It was straightforward to map this to the far more flexible NCore model of aggregators, metadata objects, and resources.

The primary metadata format used to describe resources in NSDL is a specific implementation of qualified Dublin Core called nsdl_dc. DLESE metadata is stored in two separate formats: ADN[6] and dlese_anno. DLESE provides a crosswalk from ADN to nsdl_dc, but significant information, particularly the support for DLESE's community review process provided in the dlese_anno format, is lost in the crosswalk.

Since metadata objects in NCore can support multiple independent metadata datastreams, the DLESE team simply added datastreams to support ADN and dlese_anno to the metadata object. This allows DLESE-specific processes to access the ADN and dlese_anno streams while maintaining full compatibility with all existing NCore tools and services.

## 6.2 Implementing DLESE Tools and Services

The most critical end-user functionality of DLESE is the search service. This service takes full advantage of the detailed categorization of DLESE resources represented in the ADN metadata, as well as the teaching tips, reviews, editor's summaries and other information represented in dlese_anno, to allow detailed searching and filtering. The crosswalk to nsdl_dc does not provide enough information to support this service, and DLESE's ability to use the NCore API to store and access this metadata was critical.

In fact, no change to the DLESE search service code was needed. Since the DLESE search service runs directly from index files built from the DLESE system, it was only necessary to write a process that built the index from the NDR using the API. After an initial upload of the DLESE information to the NDR and creation of the index, the search service was fully functional.

The other key DLESE tool is the Digital Collection System (DCS)[7]. This is a flexible, XML-driven cataloging tool to create and manage metadata for educational resources, as well as providing collection workflow processes. Most of the work in embedding DLESE in NSDL was in rewriting the DCS system to use the NDR API to access the DLESE ADN and dlese_anno metadata and to create and manipulate the digital objects needed to support the DLESE data model in NCore.

Since the DCS is an XML-driven system, once the changes were made to access and manipulate NCore digital objects through the NDR API, it was relatively easy to replace the existing DLESE metadata XML schema with an XML schema for nsdl_dc. At that point, the DCS became the NCS (NSDL Collection System), and the tool could be used to manipulate arbitrary collection and item metadata in the NSDL Data Repository. The NSDL project is currently in the process of replacing its former collection management system with NCS. And, as part of NCore, NCS will be available as a metadata management and cataloging tool to support any project using the NCore platform.

## 6.3 Viewing DLESE in NSDL

As it happens, the scope of the DLESE materials falls fully within the scope of NSDL. However, the aggregation and view model of NCore allows complete flexibility in the membership of resources in NSDL and in DLESE. The "DLESE view" can include only the materials uploaded and managed by DLESE, or it can also include other NSDL aggregations. The "NSDL view" can include all or only some of the DLESE collections, since aggregations can be explicitly included or excluded from the NSDL view of the library. It would even be possible to run DLESE as a completely independent digital library from NSDL within the same NCore instance of the repository.

## 7. FUTURE WORK AND PUBLIC RELEASE

As of January 2008, v1.1 of the NCore data model and API has been publicly released on SourceForge under the Educational Community License 1.0 (http://sourceforge.net/projects/nsdl-core/). The project plan calls for the existing NCore software to be released on the SourceForge site throughout 2008. These components include: updated versions of the NCore data model and API; the PHP code that implements search and browse views of the repository at nsdl.org; the ExpertVoices plug-ins for WordPress Multi-User; the NSDL Wiki plug-ins for Mediawiki; the NSDL search REST service and Lucene/Nutch-based indexing software; the NSDL Collection Service (NCS) metadata cataloging system; the OAI-PMH harvest ingest system; and a new automated OAI-PMH harvest management system.

Proposed new near-term development work on the NCore platform includes: an NCore toolkit providing Java, PHP, and Javascript tools to support the easy integration of 3$^{rd}$ party software with NCore; the ability to harvest RSS feeds, together with a system to allow individual users or organizations to register feeds for ingest into the library; and extensions to integrate NSDL with existing open-source course management systems, either Moodle, Sakai, or both.

## 8. CONCLUSION

NCore implements a flexible, extensible platform for creating a new kind of digital library that integrates the best features of traditional libraries with the collaborative tools of Web 2.0 to empower the collective creation of library materials and context by any community in any discipline. NCore has already demonstrated the ability to integrate different off-the-shelf open-source tools and to support different digital libraries. The flexible architecture and implementation of aggregations has been one key to the power and versatility of the NCore platform.

NCore provides a compelling suite of data models, services, and end-user tools combined with the proven ability to support a large, production digital library. It serves as both a model for digital library architectures and implementations and as an open-source platform on which digital library creators can build their own production systems. Finally, NCore embodies a vision of a new generation of collaborative, community-driven digital libraries that fully integrate with all the tools, infrastructure, and social and informational networks of the World Wide Web.

---

[6] http://www.dlese.org/Metadata/adn-item/index.htm

[7] http://www.dlsciences.org/tools-dls/#dcs

## 9. ACKNOWLEDGMENTS
This material is based upon work supported by the National Science Foundation under Grants No. DUE-0733600, 0424671, 0227648, and 0227888. The authors wish to gratefully acknowledge the efforts and support of the DLESE/DLS projects and development team, with particular thanks to Tamara Sumner, Michael Wright, Kathryn Ginger, Jonathan Ostwald, and John Weatherley. Thanks are also due to the entire NSDL Core Integration team at Cornell, UCAR, and Columbia. Finally, particular thanks go to James Blake, Tim Cornwell and Carl Lagoze for their contributions to this paper and the research described herein.